\documentclass[%
reprint,
superscriptaddress,
 amsmath,amssymb,
 aps,
prb,
floatfix,
]{revtex4-1}

\usepackage{fancyhdr}
\usepackage{lipsum}% just to generate text for the example
\usepackage{multirow}

\usepackage{graphicx}% Include figure files
\usepackage{dcolumn}% Align table columns on decimal point
\usepackage{bm}% bold math
\usepackage{hyperref}% add hypertext capabilities
\usepackage{color}
\hypersetup{
 colorlinks   = true, %colors links instead of ugly boxes
  urlcolor     = blue, %color for external hyperlinks
 linkcolor    = red, %color of internal links
 citecolor   = blue %color of citations
}
\usepackage{comment}
\usepackage{xfrac}
\usepackage{multirow}
\usepackage{tikz}
\usetikzlibrary{patterns}

\def\vp{\mbox{\boldmath$p$}}

\def\vq{\mbox{\boldmath$q$}}

\def\cdie{\varepsilon}
\def\simpt{{\Sigma^R_{\mathrm{imp}}}}
\def\timp{{T_{\mathrm{imp}}}}
\def\vimp{{V_{\mathrm{imp}}}}
\def\nimp{{n_{\mathrm{imp}}}}
\def\eimp{{\epsilon_{\mathrm{imp}}}}
\def\eres{{\epsilon_{\mathrm{res}}}}
\def\econe{{\epsilon_{\mathrm{D}}}}
\def\gimp{{\gamma_{\mathrm{imp}}}}
\def\teimp{{\tilde \epsilon_{\mathrm{imp}}}}
\def\sizeplotuc{.49}

\begin{document}

%\preprint{APS/123-QED}

\title{Graphene plasmons: impurities and nonlocal effects }% Force line breaks with \\

\author{Giovanni Viola}\email{giova.viola@gmail.com}
\author{Tobias Wenger}
\affiliation{Department of Microtechnology and Nanoscience (MC2), Chalmers University of Technology, S-412 96 G\"oteborg, Sweden}
\author{Jari Kinaret}
\affiliation{Department of Physics, Chalmers University of Technology, S-412 96 G\"oteborg, Sweden}
\author{Mikael Fogelstr\"om}
\affiliation{Department of Microtechnology and Nanoscience (MC2), Chalmers University of Technology, S-412 96 G\"oteborg, Sweden}

\date{Draft of \today}

%\date{\today}% 
\begin{abstract}
This work analyses how impurities and vacancies on the surface of a graphene sample affect its optical conductivity
and plasmon excitations. The disorder is analysed in the self-consistent Green's function formulation and nonlocal effects 
are fully taken into account.  It is shown that impurities modify the linear spectrum and give rise to an impurity band 
whose position and width depend on the two parameters of our model, the density and the strength of impurities. 
The presence of the impurity band strongly influences the electromagnetic response and the plasmon losses. 
Furthermore, we discuss how the impurity-band position can be obtained experimentally from the plasmon dispersion 
relation and discuss this in the context of sensing.
\end{abstract}

\pacs{Valid PACS appear here}% PACS, the Physics and Astronomy
\maketitle

%%%%%%%%%%%%%%%%%%%%%%%%%%%%%%%%%%%%%%
\section{Introduction}
%%%%%%%%%%%%%%%%%%%%%%%%%%%%%%%%%%%%%%
Plasmons are electromagnetic fields resonantly enhanced by oscillations in the charge density. Due the properties of graphene, in graphene 
plasmons exhibit low losses~\cite{Woessner2015b}, tunable optical properties~\cite{Li2008}, strong optical field 
confinement~\cite{Jablan2009,Chen2012}, and environmental sensitivity~\cite{Eriksson2013,Hu2016Jul,Wenger2017}. This makes graphene 
an attractive material for next generation technologies~\cite{Ferrari2015} in sensing~\cite{Eriksson2013,Chen2011}, photonics, 
electronics~\cite{Bonaccorso2010,Schwierz2010}, and communication~\cite{Habibpour2017}. To improve device design and performance, it is 
crucial to extend the microscopic theory of plasmons to include nonlocal effects~\cite{Thongrattanasiri2012,Christensen2014,Lundeberg2017Apr}
together with the impact of defects and impurities~\cite{Araujo201298} 
in the sample as well as chemical compounds deposited on the  surface~\cite{Haldar2016,Xuezhong2016}. Defects and impurities may be due to 
the fabrication procedure, while chemical compounds can be deposited in a controlled fashion on the surface to functionalise the graphene
substrate~\cite{Araujo201298,Ferrari2015,Haldar2016,Xuezhong2016,Kaushik2017707}. Defects and impurities are inevitably sources of losses 
that must be understood in order to make high-performance samples and devices, mainly by circumventing their loss-producing effects.  

The behaviour of plasmons in pristine graphene is by now rather well studied~\cite{Wunsch2006,Hwang2007,Ramezanali2009}. The local 
transport properties are investigated in a series of articles, 
{\it e.g.},~\cite{Ando06,Peres2006,Gusynin2006,Peres2008EPL,Stauber2008Aug,Hwang2008,DasSarma2011}, including 
effects due to the impurities, phonons, and localised charges. The nonlocal effects in the presence of impurities or adatoms have been 
considered, among others, in Refs.~\cite{Principi2013,Karimi2016,Viola2017Jun}. Phonon- and electron-electron interaction has been studied 
in Refs.~\cite{Principi2013Nov,Principi2014}.

First-principles studies have determined how crystal defects or atoms on the graphene surface influence the material
properties~\cite{Leenaerts2008,Wehling2008,Ihnatsenka2011,Gmitra2013,Zollner2016,Frank2017}. Defects are seen to give rise to new bands 
whose properties depend on the density and type of defects or adsorbates. This opens the possibility to engineer the band structure of the material. 

While first-principles studies consider relatively small graphene supercells (on the order of $10^2$ atoms), many-body techniques 
are more suitable to describe properties in $\rm{\mu m}$ size devices. In this work we include impurities in a 
self-consistent $\bf t$-matrix treatment of elastically scattering impurities, and explore how their presence modify the optical conductivity 
and plasmonic behaviour of graphene. In the microscopic model used here, as described in Sec.~\ref{sec:model}, 
the nonmagnetic impurities are described as onsite, spin-preserving potentials and treated 
self-consistently~\cite{economou2006green,Peres2006,Lofwander07,Peres2008EPL}. The nonlocal transport and 
optical properties are investigated in Sec.~\ref{sec:optical}. The optical response resembles the one obtained in the relaxation-time
approximation~\cite{Rana2008,Jablan2009} in the case of dislocations in graphene, while novel features are observed if the impurity 
band is detuned from the Dirac point. In particular, it is observed that an impurity band far from the Dirac point enhances the plasmon losses. 
Finally, we discuss how the optical response and plasmonic behaviour can characterise the impurity itself (Sec.~\ref{sec:sensing}), within our
treatment. Our work emphasizes the potential of plasmon-based sensors and of contactless
characterization of samples~\cite{Yurgens12}. 
In the following the densities of electrons and impurities are given in units of $10^{12}$/cm$^2$, the energies are measured in 
eV, 1eV=241.8THz=1239 nm and the conductance in units of $\sigma_0=e^2/(4\hbar)=6.085\times 10^{-5}$ S =(16.4 k$\Omega)^{-1}$.

\begin{figure*}
\includegraphics[width=1.01\textwidth]{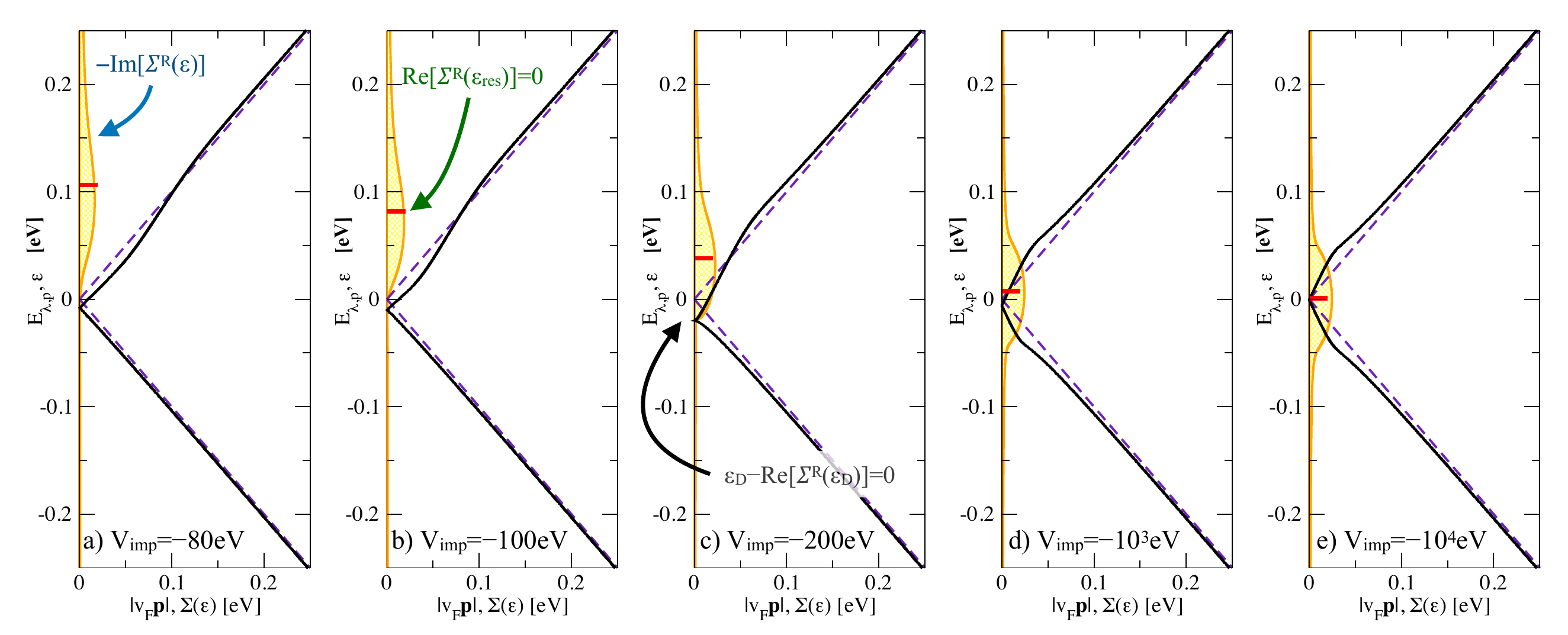}\caption{\label{fig:bandstructure} 
(Color online) The panels show how a dilute density of impurities, $\nimp=10^{-4}$, modifies the band structure of graphene. 
From a) to e) the impurity strength is increasing and the values of $\vimp$ are indicated in each panel. The full black line is the 
impurity-modified band structure, $E_{\lambda,p}=\epsilon-\mathrm{Re}\simpt(\epsilon)=0$, 
for the different impurity strengths as indicated in each panel. 
The dashed line is the pristine graphene band structure. Along each y-axis we also plot $-\mathrm{Im}\simpt(\epsilon)$ which is the 
impurity induced, frequency dependent, scattering rate which ultimately will give rise to plasmon losses. 
$-\mathrm{Im}\simpt(\epsilon)$ is centered around an impurity resonance $\eres$, marked by the thick dash on the y-axis. 
We read-off $\eres$ at $\mathrm{Re}\simpt(\eres)=0$. The final quantity we note in the figures is the shift of the Dirac cone, 
$\econe$, which is extracted at $\econe-\mathrm{Re}\simpt(\econe)=0$.}
\end{figure*} 

%%%%%%%%%%%%%%%%%%%%%%%%%%%%%%%%%%%%%%
\section{Model}\label{sec:model}
%%%%%%%%%%%%%%%%%%%%%%%%%%%%%%%%%%%%%%
Longitudinal plasmons confined at a conducting interface between two dielectrics, with relative dielectric constants 
$\varepsilon_1$ and $\varepsilon_2$, satisfy the dispersion relation \cite{Jablan2009,peres_book2017} 
\begin{equation}\label{eq:MEdisp}
\frac{(\cdie_1+\cdie_2)}{q}+\frac{i\sigma(q,\omega)}{\cdie_0\omega}=0
\end{equation} 
with the wave vector, $\vq$ ($q=|\vq|$), in the graphene plane and the angular frequency, $\omega$, of the electromagnetic field. 
Here we assume the non-retarded limit, $q \gg \sqrt{\cdie_{1,2}}\omega/c$,
as the light and the plasmons have a large momentum miss-match. An efficient coupling of light to plasmon modes is possible by
introducing a dielectric grating or coupling via evanescent light modes~\cite{peres_book2017} to overcome the mismatch. 
The nonlocal longitudinal conductivity, $\sigma(\vq,\omega)=\sigma_1(\vq,\omega)+i\sigma_2(\vq,\omega)$, 
together with the dielectric environment, encodes the plasmon properties.  
As conductors in general are lossy, $\sigma_1(q,\omega)>0$, we can read from Eq.~\eqref{eq:MEdisp} that either 
$q$ or $\omega$ is required to be complex~\cite{Principi2014} to account for these losses. Connecting to scattering 
experiments {\em e.g.} in Refs. \cite{Zhan2012,Chen2012,Fei2012}, 
$\omega$ is the frequency of the incoming light and thus real-valued which leaves the wavenumber $q=q_1+iq_2$ being a complex-valued 
quantity describing the in-plane momentum, $q_1$, and the damping, $q_2$, of the plasmons. 
For a lossless dielectric, Eq. \eqref{eq:MEdisp} reduces to the two real equations 
\begin{align}
q_1=\frac{\omega\cdie_0(\cdie_1+\cdie_2)} {\sigma_2(q_1,\omega)},
& &
\frac{q_2}{q_1}=\frac{\sigma_1(q_1,\omega)}{\frac{\partial}{\partial_{q_1}}(q_1\sigma_2(q_1,\omega))}\ ,
\label{eq:q1q2}
\end{align}
to first order in $ q_2/ q_1$. The ratio $q_2/q_1$ quantifies the plasmon losses and is called the 
{\it plasmon damping ratio}~\cite{Principi2014}. 
The non-local conductivity $\sigma(\vq,\omega)$ is evaluated from the current-current response to an external electromagnetic field within RPA   
as
\begin{widetext}
 \begin{equation}
  	\chi_{j_{\mathrm{x}}j_{\mathrm{x}}}(\vq,\omega)=
			g_sg_v\frac{ie^2v_F^2}{2} \!\!\int\! \frac{d\vp\, d\epsilon}{(2\pi)}\mathrm{Tr}\!
			\left[\sigma_{\mathrm{x}} G^R(\vp,\epsilon)\sigma_{\mathrm{x}} G^K(\vp-\vq,\epsilon-\omega) 
			       +\sigma_{\mathrm{x}} G^K(\vp,\epsilon)\sigma_{\mathrm{x}} G^A(\vp-\vq,\epsilon-\omega)\right]. \label{eq:chiIntegral}
\end{equation}
\end{widetext}
and the conductivity is then given as 
\begin{equation}
\sigma(\vq,\omega)=\frac{i}{\omega}\chi_{j_{\mathrm{x}}j_{\mathrm{x}}}(\vq,\omega).
\end{equation}
The microscopic details of the material is encoded in the Green's functions $G^R,G^A=(G^R)^\dagger,$ and $G^K$ 
and the self-energies $\Sigma^{R,A,K}$.
Here we consider a dilute ensemble of s-waves scatters, smooth on the atomic scale, included via a self-consistent $\bf t$-matrix
method~\cite{Peres2006,Ando06,economou2006green,Lofwander07,Peres2008EPL}. It is a two-parameter theory with $\vimp$, being 
the strength, and $\nimp$ the density of impurities respectively. 
The density $\nimp=N_{\mathrm{imp}}/N$, is the number of impurities, $N_{\mathrm{imp}}$, 
divided by the number of unit cells, $N$, in the crystal. We will use the Dirac approximation and in this scheme the energy
scale is set by a cut-off $\epsilon_c$ related to the band-width, we set $\epsilon_c=8.2 \mathrm{eV}$~\cite{Peres2006}.

\begin{figure}
\includegraphics[width=0.45\textwidth]{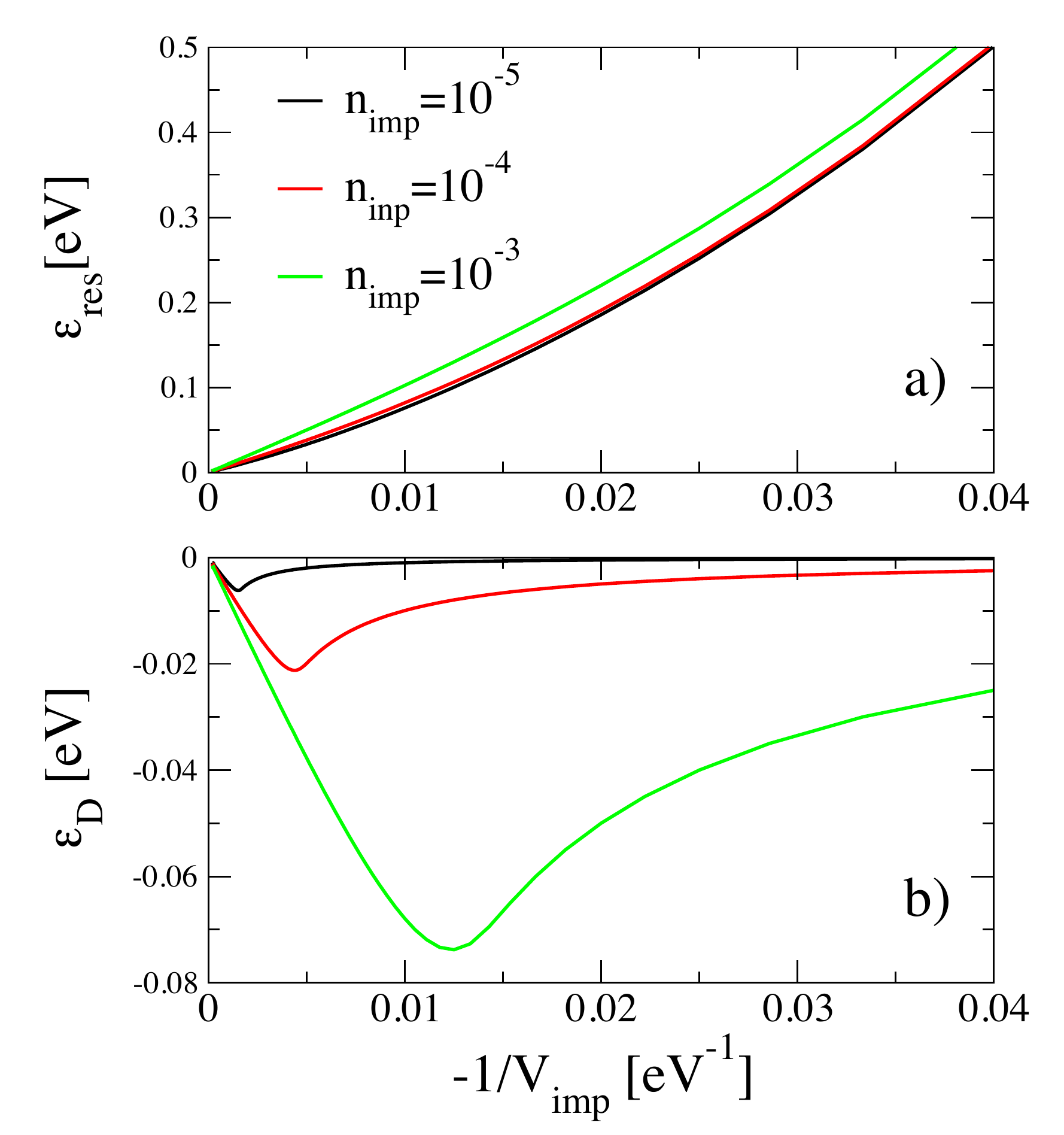}
\caption{\label{fig:diracspecs} 
(Color online) In panel (a) the location of the impurity resonance is plotted vs $-1/\vimp$ for three different impurity densities. 
In panel (b) the shift of the Dirac point $\econe$ for the same densities.}
\end{figure}

The band structure is obtained from the poles of the Green's function
\begin{equation}\label{eq:Gimp}
G^R(\vp,\epsilon)=\sum_{\lambda=\pm}
\frac{1/2}{\epsilon^R-E_{\lambda,p}-\simpt(\epsilon)}\left(\begin{smallmatrix}
1&\lambda{\rm e}^{-i\phi_p}\\  \lambda{\rm e}^{ i\phi_p}&1
\end{smallmatrix}
\right)
\end{equation}
with the self-energy
\begin{equation}
\label{eq:Sigmaimp}
\simpt(\epsilon)=\frac{\nimp\vimp}{1-\vimp\frac{1}{N}\sum_{\vp} G^R(\vp,\epsilon)}.
\end{equation}
Here $E_{\lambda,p}=\lambda v_F p$ is the single particle energy for the pristine graphene, $\lambda=\pm$, the band index, $v_F$ the 
electron momenta, $\vert\vp\vert=p$, and $\phi_p=\arg(p_x+ip_y)$. $ v_F $ is the Fermi velocity of graphene. The energies are measured 
from the Dirac point of pristine graphene.  This single Dirac-cone approximation captures the physics in the regime of interest and the 
degeneracy number $g_vg_s=4$ will be included at the end to include spin and valley degeneracy (inter-valley and spin-flip processes are omitted). 

Equation~(\ref{eq:Sigmaimp}) is derived as an average over a distribution of identical impurities as 
$\simpt(\epsilon)=\nimp \timp(\epsilon)$. The scattering off an impurity is described by the single impurity T-matrix, $\timp$.
In the Dirac-cone approximation, where the momentum sum $\frac{1}{N}\sum_{\vp} G(\vp,\epsilon)$ can be evaluated 
analytically \cite{Lofwander07}, we have 
$$\simpt(\epsilon)=\frac{\nimp}{\frac{1}{\vimp}-\frac{z}{\epsilon_c^2}\ln\lbrack\frac{-z^2}{\epsilon_c^2-z^2}\rbrack}$$
with $z=\epsilon-\simpt(\epsilon)$ and $\epsilon_c$ is a cut-off that we set to the bandwidth of the graphene band structure. 
The poles of $\simpt$ describe the well-known impurity states
which in the limit $\vimp \gg \epsilon_c$ and $\nimp$ is small (i.e. $\simpt\rightarrow 0$ so that $z\rightarrow\epsilon+i 0^+$) 
are localised states at energies
\begin{equation}
\eimp=\frac{\epsilon_c^2}{2\vimp} \ln\vert\frac{\epsilon_c}{2\vimp}\vert.
\label{eq:impres}
\end{equation} 
These low-energy impurity states are a generic feature of impurity scattering in Dirac materials \cite{Wehling2014}.
At finite impurity density, $\nimp$, the impurity state develops in to a narrow metallic band around the energy $\eimp$ 
with a width, 
\begin{equation}
\gimp\approx\sqrt{\frac{\nimp}{2 \ln\vert \frac{2\vimp}{\epsilon_c}\vert}}\epsilon_c.
\label{eq:impwidth}
\end{equation}  
$\simpt(\epsilon)$ has a simple pole structure with a complex pole $\teimp=\eimp+i \gimp$ indicating that a distribution of impurities 
induces scattering resonances rather than proper long-lived states. 

Self-consistent solution of equations~(\ref{eq:Gimp}) and (\ref{eq:Sigmaimp}) is straightforward by simple iteration. The band structure, given by 
$E_{\lambda,p}=\epsilon-\mathrm{Re}[\simpt(\epsilon)]$, is modified already for quite dilute impurity densities.  The basic results 
are shown in Fig.~\ref{fig:bandstructure}. The impurity state, $\eimp$ is modified into a resonance that we define as $\mbox{Re}\simpt(\eres)=0$.
The resonance $\eres\,$Êis shifted towards smaller energies (absolute magnitude) compared to $\eimp$. There is also an impurity dependent shift 
of the the Dirac point $E_{\lambda,p}=0=\econe-\mathrm{Re}\simpt(\econe)$. In Fig.~\ref{fig:diracspecs} we show how $\eres$ and
$\econe$ depend on the inverse of the scattering strength. The new Dirac point $\econe$ is a non-monotonous function of the inverse 
scattering strength $-1/\vimp$ and increases in magnitude with increasing impurity density. Finally, we plot $-\mathrm{Im}\simpt(\epsilon)$ 
for different scattering strength in Fig.~\ref{fig:bandstructure}. For strong scatterers $-\mathrm{Im}\simpt(\epsilon)$ has close to a 
Lorentzian lineshape around $\eres$ while for weaker scatterers $-\mathrm{Im}\simpt(\epsilon)$ has a wider spread still with weak a maximum at $\eres$. As $-\mathrm{Im}\simpt(\epsilon)$
gives an effective frequency-dependent single-particle scattering rate $\Gamma(\epsilon)=\frac{\hbar}{2 \tau(\epsilon)}$, we predict that
weak impurities will introduce plasmon losses at all frequencies along the plasmon dispersion while the losses incurred from strong impurities 
are pronounced when the plasmon mode interferes with $\eres$.      
  
In this article, $\vimp$ is chosen to be negative, giving rise to an impurity state in the conduction band, which is a common scenario suggested by 
first-principle studies~\cite{Leenaerts2008,Wehling2008,Ihnatsenka2011,Gmitra2013,Zollner2016,Frank2017}. All considerations can be repeated for 
$\vimp>0$ for which the sign of $\eres$ and $\econe$ are reversed. 

%%%%%%%%%%%%%%%%%%%%%%%%%%%%%%%%%%%%%%
\section{Optical and transport properties}\label{sec:optical}
%%%%%%%%%%%%%%%%%%%%%%%%%%%%%%%%%%%%%%
We now focus on the nonlocal graphene conductivity, $\sigma(\vq,\omega)$. 
With the knowledge of the band structure, and following the method presented in Refs.~\cite{Marsiglio1988,Viola2017Jun}, a simplified expression 
for the conductivity is obtained. The results obtained below are found to be identical if the sign of $E_F$ and $\vimp$ are reversed, due to the particle-hole symmetry. 

% fig 3
\begin{figure}[b]
\includegraphics[width=\sizeplotuc\textwidth]{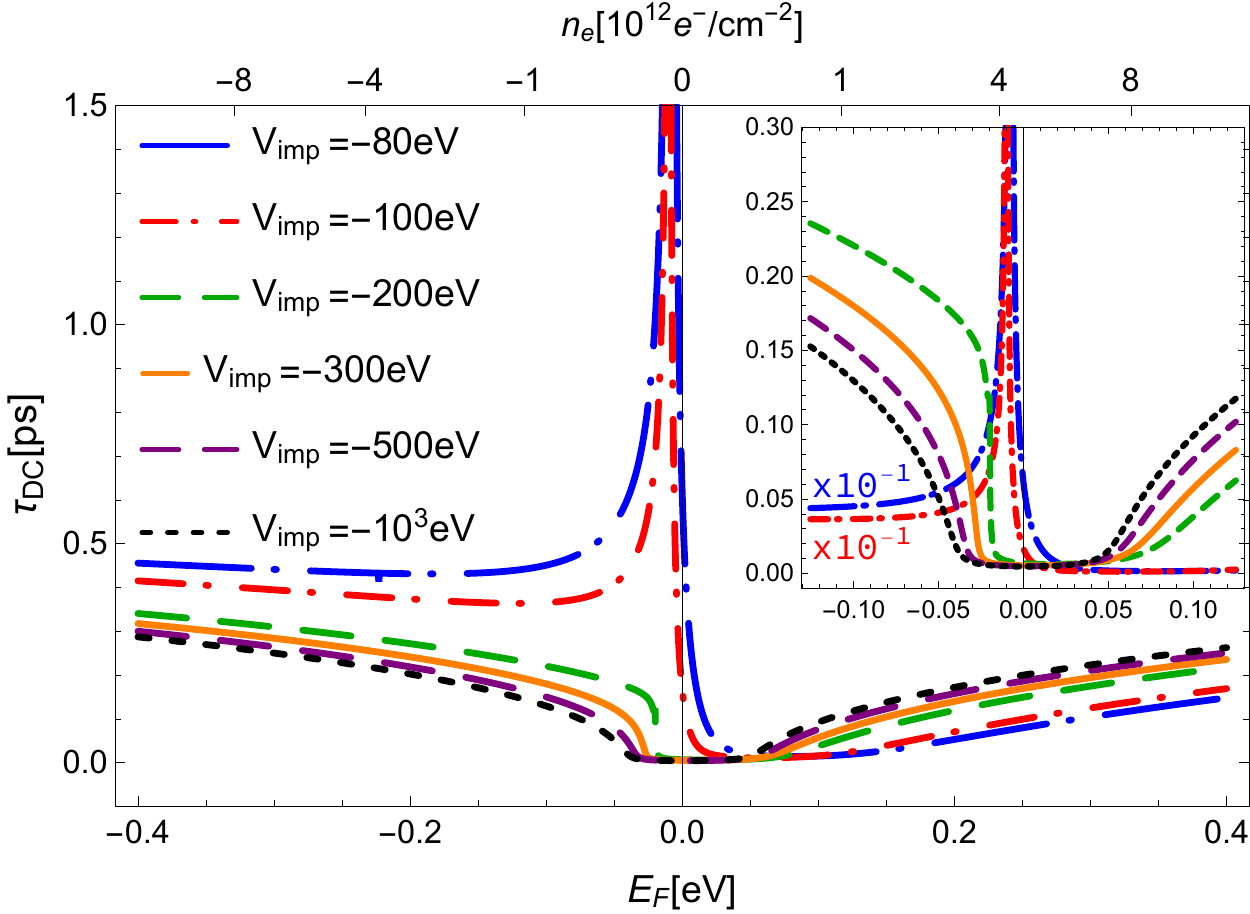}
\caption{\label{fig:taoDC} 
(Color online) The relaxation time induced by impurities obtained from equation~\eqref{eq:tauDC} as function of the Fermi energy.  
In the plot, the density of impurities is $\nimp=10^{-4}$, the relaxation time roughly scales as $1/\nimp$. In the inset a zoom for 
small value of $\vert E_F\vert$. Different colors and marks correspond to different value of $\vimp$ as indicated by the legend.}
\end{figure}

To start out we consider the impact of disorder on the local conductivity $\sigma(\vq=0,\omega)$. This has been investigated 
in detail (see, {\it e.g.}, Refs.~\cite{Ando06,Peres2006,Peres2008EPL,Hwang2008}). We revisit it nevertheless briefly with attention on how impurities 
modify the DC conductivity. The impurity contribution to the transport scattering time ($\tau_{\mathrm{DC}}(E_F)$) is given by the 
relation~\cite{DasSarma2011} 
\begin{align}
\frac{1}{\tau_{\mathrm{DC}}(E_F)}=\frac{e^2v^2_F}{2}\frac{\rho(E_F)}{\sigma_1(\vq=0,\omega\mapsto 0)}
\label{eq:tauDC}
\end{align}
in the limit  $T\rightarrow0$. The competition between the density of states and the DC conductivity gives rise to a non-trivial relation between
Fermi energy and transport scattering time as shown in Fig.~\ref{fig:taoDC}. The relaxation time is non-symmetric in $E_F$ for small $\vimp$ 
and a symmetric behaviour is recovered for large $\vimp$. Fig.~\ref{fig:taoDC} suggests that a chemical potential with the same sign as the more
common impurity strength may increase $\tau_{\mathrm{DC}}(E_F)$~\footnote{
According to Ref.~\cite{Principi2013}, impurity effects should be a
dominant relaxation process in graphene samples. This supports the statement in the main text that having sign$(E_F)=$sign$(\vimp)$ can
improve DC transport properties. To finally confirm this statement, the effect of other sources of relaxation (e.g. phonon and electron-electron interactions) should also be included at a self-consistent level \cite{Karimi2016}.
}.  
For $\vert E_F\vert>0.1\, \mathrm{eV}$ temperature effects on $\tau_{\mathrm{DC}}(E_F)$ are on order of $ 1\%$ or smaller up to 150K. 
The scattering time scales with the density of impurities as $\tau^{-1}_{\mathrm{DC}}(E_F)\propto\nimp$ as expected. In the 
following we focus mostly on the impurity density order of $\nimp\propto10^{-4}$. This density of impurities gives a relaxation time that is 
suitable for plasmonic application in the THz regime. 

%-----------
% fig 4
\begin{figure}[t]
\includegraphics[width=\sizeplotuc\textwidth]{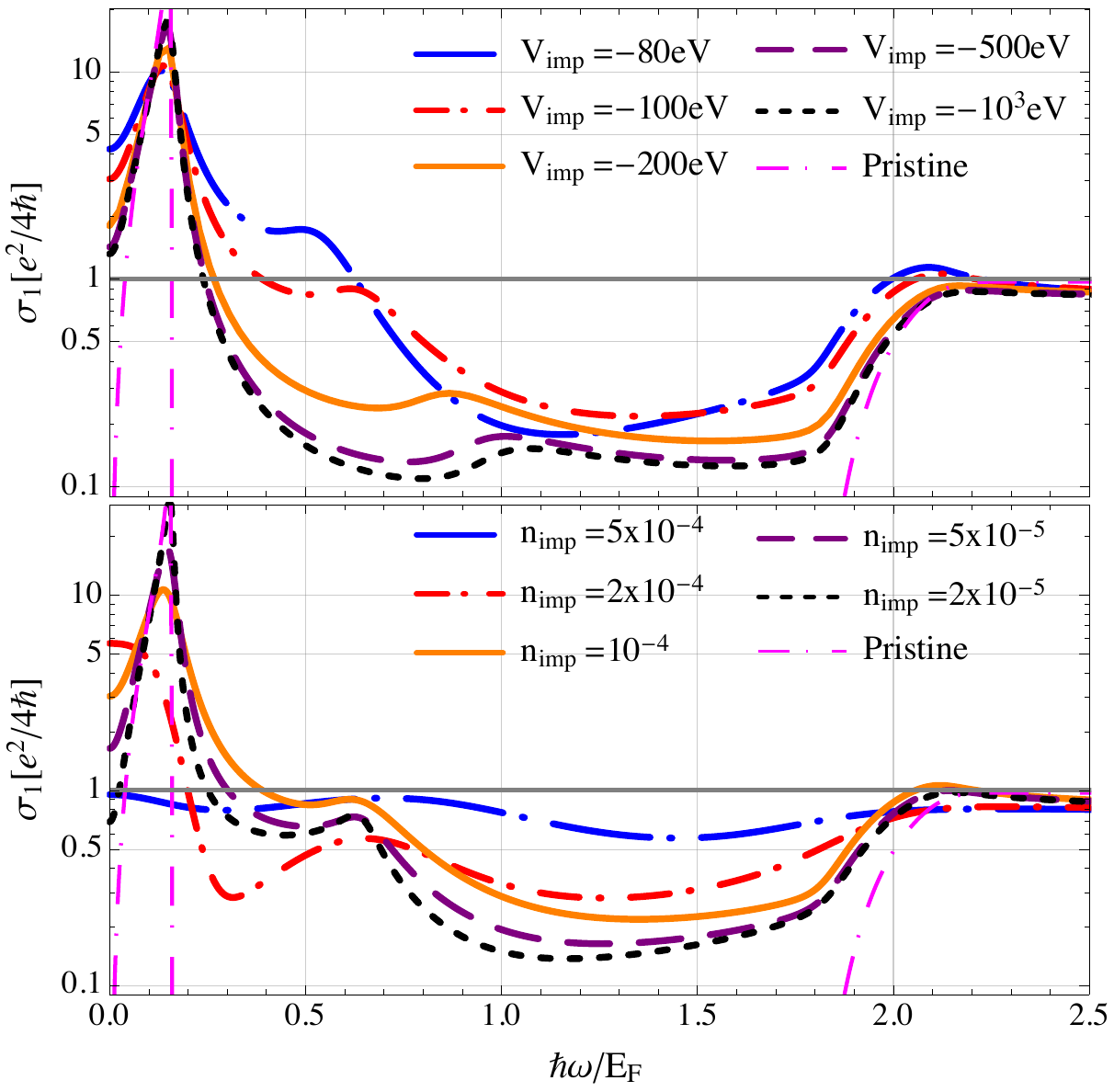}\\
\caption{\label{fig:sigma1vimp} 
(Color online) Real part of the nonlocal conductivity $\sigma_1(\vq,\omega)$ as function of the frequency. In the top panel the impurity density is set to $\nimp=10^{-4}$ and different lines correspond to different values of impurity strength $\vimp$. The fine magenta line is the pristine graphene case. Impurities induce peaks in the losses at frequency $0.4E_F\lesssim\hbar\omega\lesssim E_F$ for the values of $\vimp$ considered in the figure. The position of the peak depends on  $\vimp$, and  increasing the strength induces a blue shift of the feature. The lower plot shows the influence of the density of impurity for $\vimp=-100\, \mathrm{eV}$.  
 Here, $\sigma_1(\vq,\omega)$ is computed for  $T=30$K, $E_F=0.2\, \mathrm{eV}$ and $\vert q\vert=2\pi/\lambda$ with $\lambda=130$nm $ (q/k_F\simeq 0.15$)~\cite{Wenger2017}.    }
\end{figure}
%---------------

Now we turn to the consequence of the impurity band on the nonlocal conductivity. According to the Fermi Golden rule, the lossy part of the
conductivity ($\sigma_1$) gives the possibility of the electromagnetic field to release energy to the carriers in graphene by exciting electron-hole
pairs~\cite{peres_book2017}. In pristine graphene, there is a triangle in the ($\omega,\vq$)-plane, given by 
$\hbar v_Fq<\hbar\omega<2\omega- \hbar v_Fq$, 
where absorption is forbidden, i.e., $\sigma_1(\vq,\omega)=0$ at $T=0$. Absorption is allowed outside this Pauli-blocked triangle, i.e. for 
$\hbar\omega\leq\hbar v_F k $ and $2E_F-\hbar v_F k \leq\hbar\omega$. The absorption occurs in intraband and interband transitions
respectively~\cite{Wunsch2006,Hwang2007,peres_book2017}. It has also been shown that for pristine graphene the response depends only
on one energy scale, the Fermi energy, which scales all energies ($\omega,~E_{\lambda,p}$ and $k_BT$)~\cite{peres_book2017}. 

 %----------------------
%fig5---------
\begin{figure*}[ht!]
\includegraphics[width=.99\textwidth]{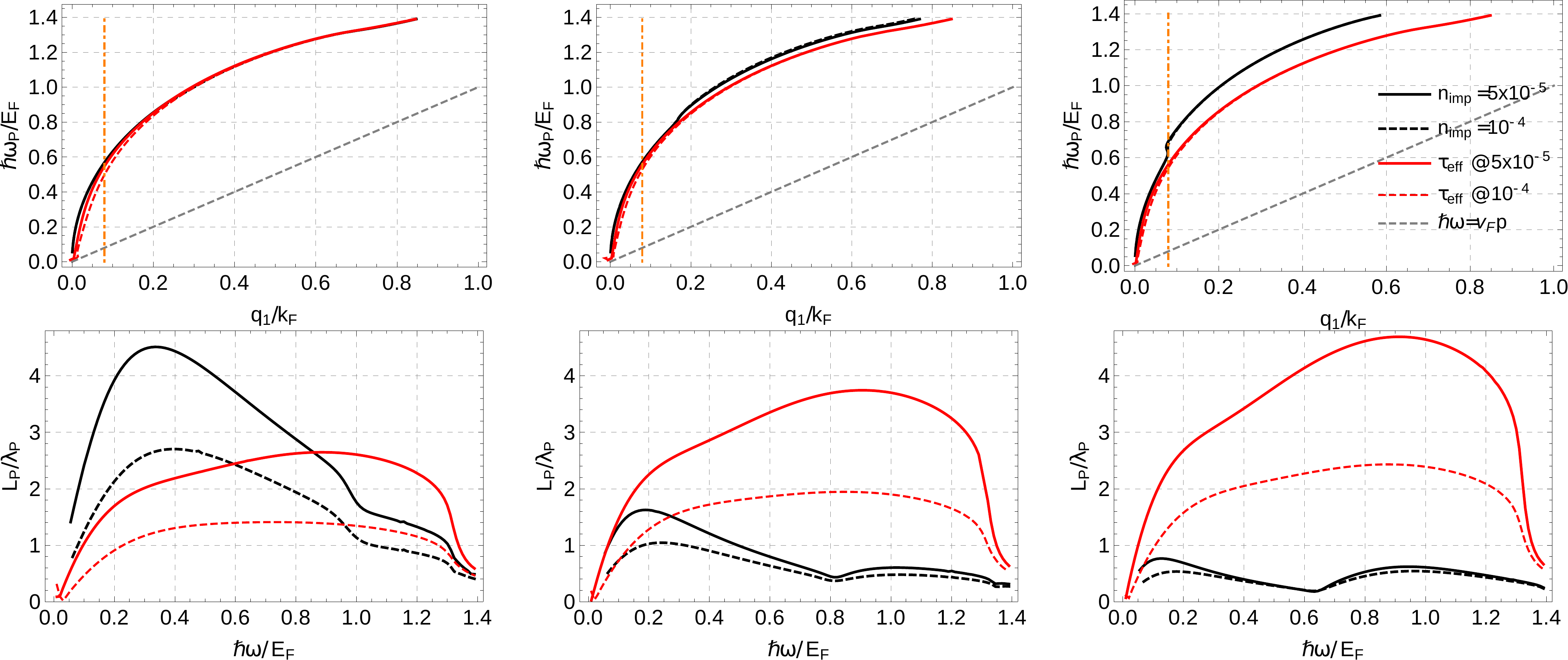}\caption{\label{fig:disprelation} 
The plasmon dispersion relation, $\hbar\omega_P(q_1)$, and propagation length, $L_P/\lambda_P=q_1/(4\pi q_2)$ 
are plotted for different values of impurity strength and density. The impurity strength from the left to the right takes the values 
$\vimp=-1000,~-100,~-60\, \mathrm{eV},$. The impurity density $\nimp=10^{-4}$  and $\nimp=5\times10^{-5}$ are 
the dashed and full black lines respectively.
$q_{1,2}$ were computed according to Eq.~\eqref{eq:q1q2}. The plasmon dispersion is given in units of $E_F/\hbar$ 
(here $E_F=0.4\, \mathrm{eV},$). The propagation length is scaled by the plasmon wavelength.  
The results are compared with relaxation-time approximation results evaluated at the marked scattering times $\tau_{\mathrm{eff}}$. 
$\tau_{\mathrm{eff}}$ is chosen so that for each pair $(\vimp,\nimp)$ the DC relaxation time is extracted 
from relation~\eqref{eq:tauDC} and used to compute the dispersion relations according to Ref.~\onlinecite{Jablan2009}. 
In the top row: the grey dashed line is the single particle continuum $\hbar \omega_P =\hbar v_F q_1$, and the vertical dash-dotted 
orange line correspond to the wavelength of $\lambda=130$nm for $E_F=0.4\, \mathrm{eV}$.
The plots show a strong non trivial behaviour as a function of the parameters $\nimp$ and $\vimp$. The dip in the 
ratio $L_P/\lambda_p$ occurs at $\hbar\omega\approx|E_F-\eres|$ where losses are enhanced.}
\end{figure*} 

In Fig.~\ref{fig:sigma1vimp} we plot $\sigma_1(\vq,\omega)$ at $q/k_F=0.15$ as function of frequency $\omega$. 
The impurity specific features we find are interband processes corresponding to transitions between the impurity band around $\eres$ 
and the states around the Fermi energy.  The transitions generate an extra peak in $\sigma_1(\vq,\omega)$ at frequencies  
$\hbar\omega\simeq\vert\eres-E_F\vert$ added to the main peak at $\hbar\omega/E_F=0.15$ as seen in Fig.~\ref{fig:sigma1vimp}.
These transitions are single particles excitations~\footnote{
For $\eres-E_F<0$, the transitions between single particle states 
in the impurity band and states above the Fermi energy contribute to an enhancement of losses at this frequency 
(while for $\eres-E_F>0$ the relevant transitions are from the Fermi surface to the impurity level).
}. In the lower panel of Fig.~\ref{fig:sigma1vimp} an extended range of density of impurities is considered. 
The extra peak in $\sigma_1(\vq,\omega)$ follows $\eres$ and is actually a band-like area in the 
$(\omega,\vq)$-plane around $\omega\simeq\vert E_F-\eres\vert$ where there are increased losses, 
independent of $q$. The width of this stripe is given mainly by $-\mathrm{Im}\simpt(\epsilon)$. Temperature effects are 
also important but only at high temperatures. The lower panel of 
Fig. \ref{fig:sigma1vimp} shows how the density of impurities, and hence $-\mathrm{Im}\simpt(\epsilon)$, affects 
transport properties. Since $\simpt\propto\nimp$, increasing the impurity density all features in the conductivity are broadened 
and $\sigma(\vq,\omega)$ tends to $\sigma_0=\mathrm{e}^2/4\hbar$. For a given $\nimp$, weaker impurities have a stronger 
effect on AC transport. For the range of parameters explored, the impurity induced peak emerges distinctly 
above the background at $T<200$K. The new features in $\sigma_1$ reflect in a non-monotonic behaviour of $\sigma_2$, according 
to the Kramers-Kronig relations~\cite{vignale}. Similar features are given by the impurity states in the presence of adatoms on the 
graphene surface~\cite{Viola2017Jun}. The work presented here confirms that the main features survive in a self-consistent 
treatment of impurities.

There is a minor mismatch between the computed position of the peak in $\sigma_1$ and the estimate $\vert E_F-\eres\vert$. 
This is due to the impurity-induced energy renormalisation $\mathrm{Re}\simpt(\epsilon)$ which modifies the band structure.
This means that if the real part of the self energy is omitted there are inaccuracies in the values of conductivity and, finally, 
of resonances and losses of the circuit that embed the graphene sample. In this study it is found that the error in the resonance 
frequency does not exceed $5\%$ for $2\times10^{-6}<\nimp <2\times10^{-4}$.  

\begin{figure*}[ht!]
\includegraphics[width=0.99\textwidth]{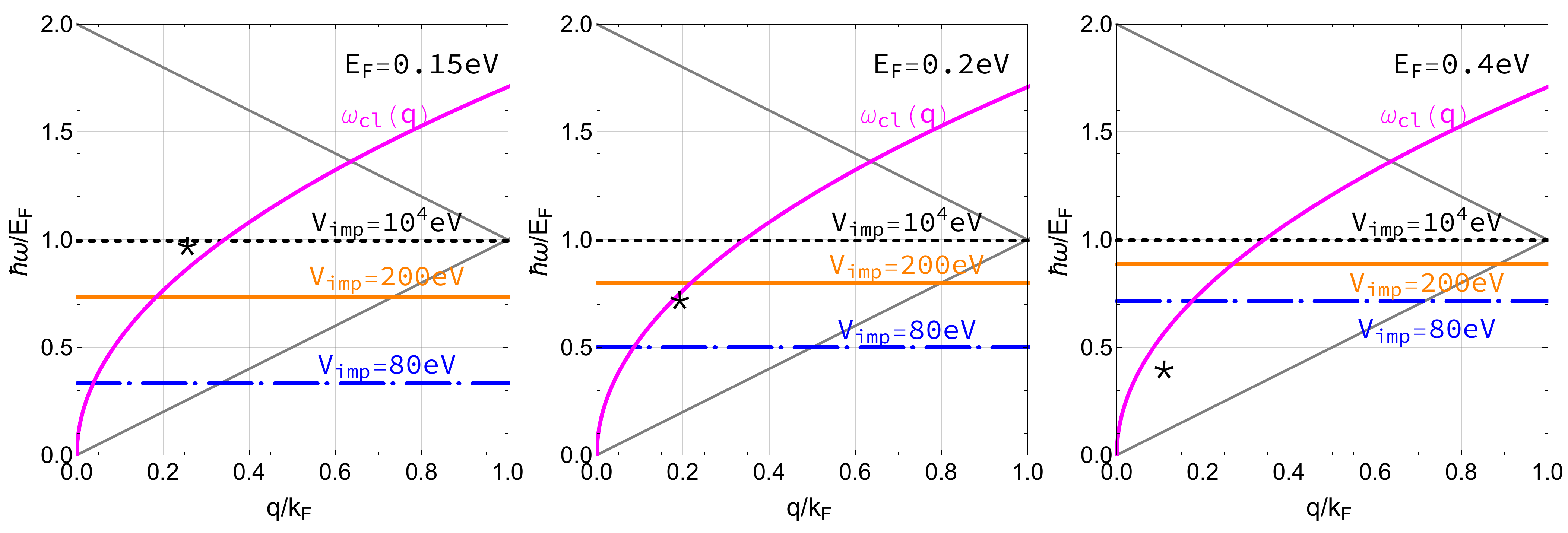}\\
\caption{\label{fig:threeEf} 
(Color online) Plasmon dispersion, in a long wavelength approximation (purple solid line), together with the impurity transitions for different impurity strengths shown for three different chemical potentials. The key insight in the sensing scheme described in section \ref{sec:sensing} is that a fixed light frequency and grating may effectively probe different regions of the plasmon dispersion by use of the gate tunability. Furthermore, this allows the effect of different impurities to be probed since the effect of impurities is largest when the impurity transitions are resonant with the graphene plasmon. The * in the figures mark the point in parameter space that is probed with a light frequency of $0.15$ eV and a periodicity of $130$ nm (same as in Sec.~\ref{sec:sensing}).}
\end{figure*}

Now we turn to how the presence of an impurity band affects the plasmonic properties of graphene. As the impurities influence both the dissipative ($\sigma_1$) 
and kinetic ($\sigma_2$) part of the conductance they also affect both the plasmon dispersion relation $\omega(q_1)$ and the 
propagation length $L_P= 1/(2q_2)$~\footnote{
In this work the propagation length is defined according to Ref.~\cite{peres_book2017} 
as the distance that the plasmon propagates until the intensity is reduced by a factor 1/e$\simeq 0.37$.
}. 
In Fig.~\ref{fig:disprelation} the top row shows the plasmon dispersion relation $\omega(q_1)$. 
The bottom row in Fig.~\ref{fig:disprelation} presents the propagation length 
$L_P(\omega)$ in units of the plasmon wavelength $\lambda_P=2\pi/q_1$, for the same values of $\vimp$ and $\nimp$ as in the panel directly
above. In each column two values of impurity density are shown, $\nimp=5\times10^{-5},\,10^{-4}$, and the impurity strength changes 
with the column.  The dispersion relation obtained from the impurity-doped graphene is compared with results from a relaxation time 
approximation~\cite{Mermin1970,Jablan2009} using the DC relaxation time value computed according to the finite temperature equivalent of 
Eq.~\eqref{eq:tauDC}~\cite{DasSarma2011}. The analysis of the losses shows a disagreement between the two approaches as was observed in
Refs.~\cite{Hwang2008,Principi2013,Tassin2013} and here confirmed in a self-consistent t-matrix model. The effects of the 
impurities are fully considered also in the evaluation of the dispersion relation $\omega(q_1)$. As seen in the figure there
is quite a discrepancy between the relaxation time approximation and our impurity model. The impurity model shows a 
clear signature of the impurity resonance as the frequency is swept. This is particularly clear for the strong scattering case. For weaker scatterers
we also see as structure in the propagation length at $\eres$ as well as a shift in the plasmon dispersion at $\hbar \omega=\eres$. 
While the relaxation time approximation is able to show damping, it is clear that impurities in the self consistent model contains features that are not captured at all in the relaxation time approximation.

To analyse the dispersion relation in more detail we use a rather large value of the chemical potential $E_F=0.4\, \mathrm{eV}$ 
and a temperature of $30$ K. The results are shown in Fig.~\ref{fig:disprelation}. The purpose of this choice is to enhance the visibility 
of the effects of the impurity band. Thanks to the approximate scale invariance of the system, the main features remain valid also for 
smaller values of the Fermi energy. However, the position of the impurity resonance needs to be wisely rescaled and one must keep in mind 
that line shape of $-\mathrm{Im}\simpt(\epsilon)$ becomes broader the further away $\eres$ is from the Dirac point. 
The left panel of Fig.~\ref{fig:disprelation} shows the case of strong impurities
($\vimp=-10^3\, \mathrm{eV}$), this may  represent a graphene lattice with dislocations or holes in it. The impurity resonance is expected 
around $\eres\simeq 0.02,\,0.04\, \mathrm{eV}$ for the densities used in the plot $\nimp=5\times10^{-5},\,10^{-4}$ (full and dashed lines) respectively. 
The signature of $\eres$ is a marked drop in the propagation length, $L_p/\lambda_p$, at $\hbar\omega/E_F\simeq0.98(0.96)$ for 
$\nimp=5\times10^{-5}(10^{-4})$.  
This brings us to a first conclusion: holes and  dislocations in the graphene crystal reduce the bandwidth of the plasmons to 
$\hbar\omega<E_F$ from the range $\hbar\omega<1.3E_F$ that the relaxation time approximation approach suggests.~\cite{Jablan2009,Wenger2017}
We do not consider effects of phonons to underline pure impurity effects. According to the literature an optical phonon introduces an 
extra bound  to the working frequency of graphene plasmons to $\hbar\omega<0.2\, \mathrm{eV}$.~\cite{Ishikawa2006Agu,Jablan2009} 
In Fig.~\ref{fig:disprelation}, the second and third column, present the dispersion relation for impurities bands that lie away from the Dirac point. 
The column in the middle display the case when $\vimp=-100\, \mathrm{eV}$ and corresponds to an impurity band around 
$\eres\simeq0.082\, \mathrm{eV}$ and $\eres\simeq0.093\, \mathrm{eV}$  for $\nimp=5\times10^{-5}$ and $\nimp=10^{-4}$ respectively. 
A clear increase in over-all losses appear and  close to $\hbar\omega\simeq E_F -\eres$ we see a signature of $\eres$ as dip in $L_p$. 
In the right column, the dispersion relation for impurities with strength $\vimp=-60\, \mathrm{eV}$. 
Now the impurity band is even higher up in the conductance band compared to the case with $\vimp=-100\, \mathrm{eV}$ and we find 
$\eres=0.15\, \mathrm{eV}$ and $\eres=0.16\, \mathrm{eV}$ for the two densities. This reflects the even more lossy conductance 
and $L_p/\lambda_p\lesssim 1$ 
for all frequencies.
For large impurity densities, $\nimp>10^{-3}$, the longitudinal plasmons appear to be overdamped according to 
Eq.~\eqref{eq:q1q2} and it may not be appropriate to speak about modes. This suggests one obvious reason why graphene of too low quality 
is not suitable for (longitudinal) plasmonic applications. 

\begin{figure*}[ht]
\includegraphics[width=0.99\textwidth]{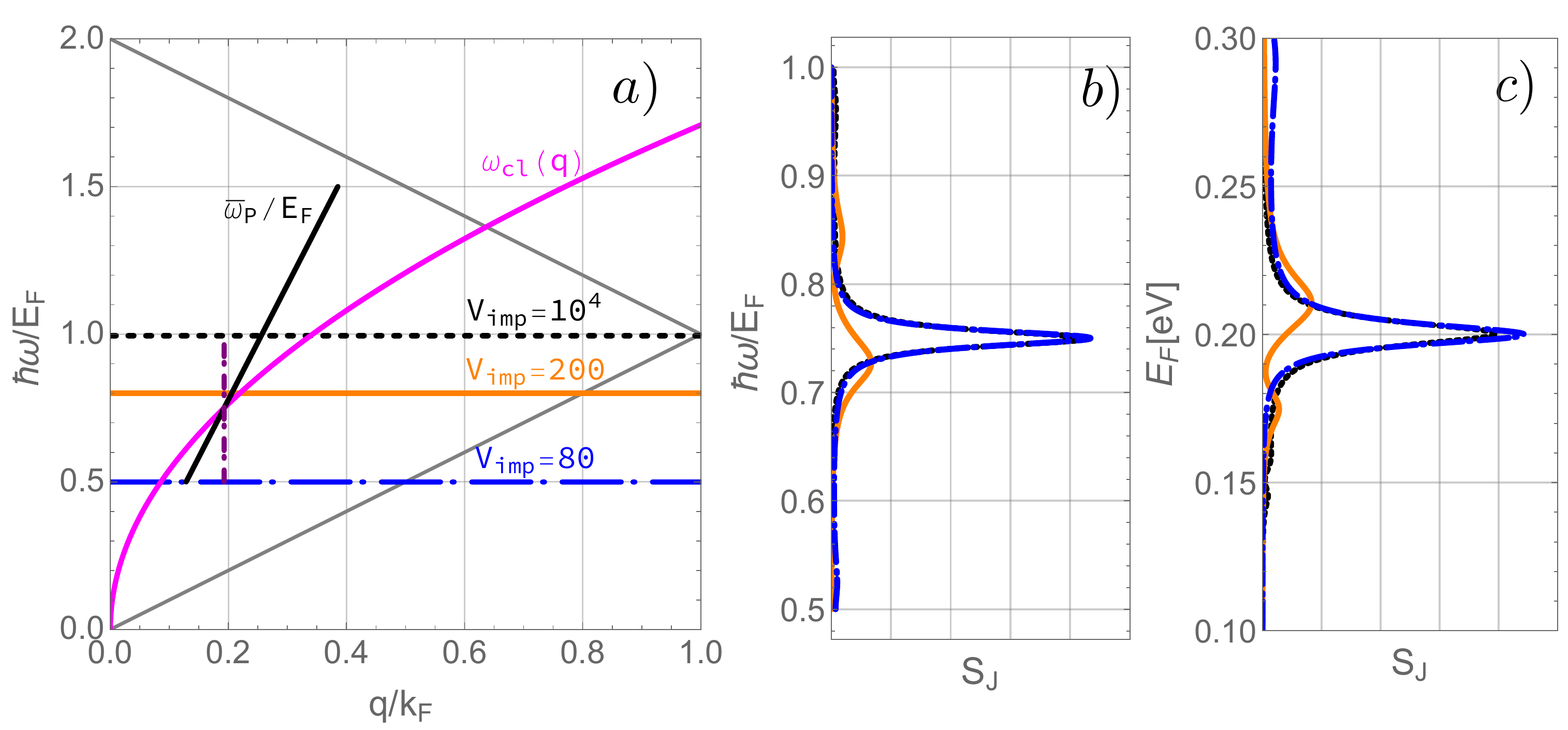}\\
\caption{\label{fig:Snew} 
(Color online) A simple model is used to gain insight in the interplay between plasmons and impurity levels. The plasmons are here considered in a long wavelength approximation giving rise to a $\sqrt{q}$ behaviour of the dispersion (purple solid line). The damping is modeled using a Drude conductivity with an energy dependent scattering time $\tau^{-1}=\tau_0^{-1}+\tau(E)^{-1}$ where $\tau(E)^{-1}$ has a Lorentzian shape around the impurity level. a) Plasmon dispersion relation and impurity transitions shown for $E_F=0.2$ (same as middle panel in Fig.~\ref{fig:threeEf}). The black solid line shows the region probed when changing the Fermi energy from $0.1$ eV to $0.3$ eV for incident light frequency $0.15$ eV and periodicity $130$ nm. To highlight the difference between working with a fixed Fermi energy and tunable Fermi energy, the vertical dashed purple line shows a cut in frequency (and $E_F=0.2$ eV is fixed) which can be obtained by tuning the incidence frequency. b) The loss function obtained by plotting the loss function following the vertical cut in panel a). The different colors correspond to having the impurity level at the locations indicated by the horizontal lines in panel a). c). The loss function obtained for a fixed light frequency and a fixed grating when changing the Fermi energy, i.e., along the black solid line in panel a).}
\end{figure*}

The comparison between temperature and finite momenta losses, considered in~\cite{Wenger2016}, and impurity losses reveals that the last 
are dominating up to room temperature for $\omega/E_F<1.2$ and for $\nimp>5\times10^{-5}$. At lower density $\nimp\simeq\times10^{-5}$ 
the two source of losses are comparable in the range of frequency  $1<\omega/E_F<1.4$ and impurity losses are dominant at lower frequency.

%%%%%%%%%%%%%%%%%%%%%%%%%%%%%%%%%%%%%%
\section{Plasmons as chemical sensing tools}\label{sec:sensing}
%%%%%%%%%%%%%%%%%%%%%%%%%%%%%%%%%%%%%%
In the previous section, the effects of impurities on the optical conductivity of graphene and the graphene plasmon resonance were investigated. The effect of the impurities 
on the plasmons is most pronounced when the plasmon and the impurity transitions are in resonance with each other. This property constitutes an indirect way to transduce optical energy into the impurity band around $\eres(\vimp,\nimp)$, in a way that can be specific 
for a given species of molecules on the surface. One of the main advantages of graphene plasmons is given by the tunability of the optical 
properties in graphene. By adjusting the Femi energy in graphene, the plasmon resonance can be tuned~\cite{Li2008} in and out of resonance 
with the impurity transition so that probing the impurity level position becomes possible. This flexibility could be relevant to overcome the constraints introduced by the structure that allows to couple light and plasmons~\cite{peres_book2017}. Below, a measurement
protocol to reconstruct the impurity resonance position is proposed.

The impedance matching required to couple graphene plasmons with light can be achieved by coupling via STM tips~\cite{Chen2012,Fei2012} or by introducing a periodic structure, either a dielectric grating~\cite{Zhu2013}, or a patterned graphene sheet~\cite{peres_book2017}.
The periodicity fixes the value of the wave vector $\vq$ to couple the electric field to the plasmons, but also reduces the phase space that can be explored, hence the
information that can be collected. There are still two degrees of freedom which can be used: the Fermi energy, accessible by gating the graphene,
and varying the incident light frequency. In this article we explore the first possibility, while the second has been discussed in Ref.~\cite{Viola2017Jun}. 
The structure of the current-current loss function~\cite{peres_book2017}
\begin{eqnarray}
S_{j_\mathrm{x}}(q,\omega)
=  \frac{\omega \sigma_1(q,\omega)}{\vert 1+\frac{iqe^2\sigma(q,\omega)}{\omega\varepsilon_0(\varepsilon_1+\varepsilon_2)}\vert^2} 
\end{eqnarray}
indicates where one may deposit energy in the sample, for instance via electromagnetic radiation. The strongest response is found at sharp maxima in
$S_{j_\mathrm{x}}(q,\omega)$, and these peaks coincide with the plasmon dispersion. This property has been used in 
Ref.~\cite{Yan2013} to map out the dispersion relation of plasmons in graphene. In this article, we take advantage of the new structures arising in the loss function due to impurity scattering. We use these features to determine the parameter $\vimp$, the value of which represents a certain type of impurity on the graphene surface, and $\nimp$ which is the density of impurities.

Before going to the full numerical results, it is useful to consider a simple model for plasmons and the impurities in order to gain insight into the sensing properties. Fig.~\ref{fig:threeEf} shows the plasmon dispersion, using the long-wavelength approximation, together with the impurity transitions. This is shown for three different Fermi energies and illustrates how the impurity transitions shift with respect to the plasmon dispersion. The idea is to use this property to distinguish between different positions of the impurity level by observing the graphene plasmon. Indeed, it was shown in Fig.~\ref{fig:disprelation} that impurities can severely affect the graphene plasmons by inducing large damping. The left panel in Fig.~\ref{fig:Snew} shows the plasmon dispersion together with the line (black solid line) in parameter space that is probed when varying the Fermi energy from $0.1$ eV to $0.3$ eV. More specifically, the black solid line is obtained by calculating the point in parameter space that is probed for the specific fixed frequency ($0.15$ eV) and periodic structure (periodicity $130$ nm) considered. This point moves in the left panel since what is plotted is energy and wavenumber divided with $E_F$ and $k_F$ and these quantities are tuned. The vertical purple dashed line shows for comparison a vertical cut made by changing the incident light frequency. These two cuts represent two different ways of probing plasmons in a fixed periodic environment. The loss function obtained for the two cuts are shown in the middle and right panels of the figure. In both panels, the plasmon peak is severely affected by the impurity transition that is resonant with the plasmon at the point in parameter space where it is being probed.

Having obtained some insight from the simple model above, we now consider the full numerical model and the sensing protocol is suggested as follows. First the periodicity in space $\bar\lambda=2\pi/\bar q$ of the electric field is fixed by the periodic structure. This is needed to couple plasmons with the incident light. Quantities with a bar on top, e.g. $\bar\lambda$, remain fixed throughout the procedure which is in contrast to the Fermi energy that will be tuned. The light frequency considered is on resonance with the plasmon 
at a given Fermi energy ($\bar E_F$) for a given reference sample, for example, with dislocations, i.e. $\vimp\mapsto \infty$. The incident light frequency is fixed for the rest of the procedure and denoted with $\bar\omega$. The behavior of the loss function while tuning the Fermi energy is recorded to be used as the reference for the next set of measurements. These measurements on further samples with unknown impurity types are done again by recording the loss function values for the same light frequency while varying the Fermi energy as before.

\begin{figure}[t]
\includegraphics[width=\sizeplotuc\textwidth]{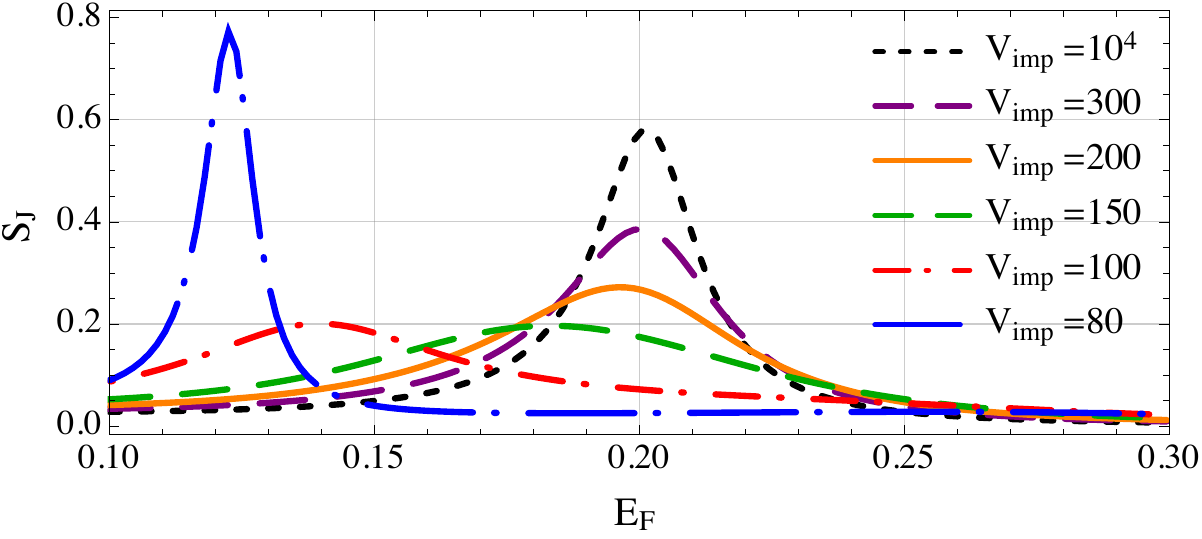}\\
\caption{\label{fig:sucts} 
(Color online) Loss function plotted for different impurity strengths as function of the Fermi energy. The impurity density $\nimp\,$ is $5\times10^{-5}$ 
for all curves. The dashed black line corresponds 
to the calibrating measurement on known sample with only dislocations $\vimp\mapsto\infty$. For the other lines the impurity 
level is moved higher in the conduction band, which can be attributed to unique scattering properties of previously tested chemical compounds. 
The impurity bands are located at $\eres=0.046\, \mathrm{eV}$ for $\vimp=200\, \mathrm{eV}$, $\eres=0.056\, \mathrm{eV}$ for $\vimp=150\, \mathrm{eV}$, 
$\eres=0.082\, \mathrm{eV}$ for $\vimp=100\, \mathrm{eV}$ and, $\eres=0.1\, \mathrm{eV}$ for $\vimp=80\, \mathrm{eV}$. In this plot the loss function is evaluated at room temperature and an additional relaxation time of $0.3$ ps is introduced to take into account additional relaxation processes.}
\end{figure}

Fig.~\ref{fig:sucts} shows the loss function for the full numerical results for various impurity strengths at room temperature. The graphene sample considered in
Fig.~\ref{fig:sucts} is assumed to be in vacuum, i.e. suspended. Results for graphene on a substrate are qualitatively similar, however, the plasmon dispersion is somewhat shifted due to the presence of the substrate. The wavelength of the electric field periodicity chosen here is equal to 
$\bar\lambda=130$nm and this corresponds to a plasmon frequency of $\omega_P\simeq0.15\, \mathrm{eV}$ for $E_F=0.2\, \mathrm{eV}$ 
of the reference sample with dilute densities of impurities $\nimp=5\times 10^{-5}$ and scattering strength of $\vimp=10^4$eV. In Fig.~\ref{fig:sucts}, the plasmon-peak shapes and positions are strongly influenced by the impurity transitions, 
i.e., the impurity strength $\vimp$. We can from the calculated loss function extract the corresponding impurity properties, which 
may be a fingerprint of a given chemical compound, assuming known impurity density. While the strength of the impurity scattering shifts the peak position, the density mostly controls the width of the peak. Further analysis is needed improve the protocol in order to extract both quantities. In this work we have focused on demonstrating the possibility of sensing microscopic degrees of freedom as well as using the unique possibility given by the tunability of graphene plasmons.

%---------------
% fig 6
%---------------

%%%%%%%%%%%%%%%%%%%%%%%%%%%%%%%%%%%%%%
\section{Conclusion}\label{sec:conc}
%%%%%%%%%%%%%%%%%%%%%%%%%%%%%%%%%%%%%%
In this work, a microscopic model of plasmons that considers together impurities~\cite{Principi2013} and nonlocal
effects~\cite{Lundeberg2017Apr,Wenger2016}, has been developed  and analysed. The present work also contributes 
to shaping a full microscopic picture of the plasmon in graphene~\cite{Karimi2016,peres_book2017} with the long term 
aim to develop further the design of plasmonic devices.

The first-principles computations~\cite{Leenaerts2008,Wehling2008,Ihnatsenka2011,Gmitra2013,Zollner2016,Frank2017} suggest 
that impurities on the graphene surface introduce an almost flat impurity band close to the Dirac point, whose width and position 
in energy depend on the type and density of impurities. This is the main feature that is newly included in theoretical description 
in this work. An impurity model based on the $\bf t$-matrix formalism in the Green's function
framework~\cite{Ando06,Gusynin2006,Peres2006,Peres2008EPL,Stauber2008Aug} is developed and analysed. The band structure 
of the model shows an impurity band similar to first-principles results, Sec. \ref{sec:model}. The model has two parameters,  
density and impurity strength, which control the band structure of graphene with impurities. An approximate mapping between 
the DFT band structure and the theoretical impurity model has been demonstrated to be possible. 

It is found here that induced impurity losses have large effects on the optical properties: for frequencies in resonance with 
the transition from the impurity band to above the chemical potential, a substantial increase of the losses is obtained, 
Sec.~\ref{sec:optical}. This has relevant effects on the dispersion relation for the plasmons. Both the relation $\omega(q)$ 
and the plasmon damping rate depend on the impurity type. The impurity effect emerges also in the optical conductivity 
$\sigma(\vq,\omega)$, and an enhancement of the losses, $\sigma_1(\vq,\omega)$, is observed when the incident light 
frequency is in resonance with the possible transitions involving the impurity states. 

The possibility to identify the type of surface impurities from their effect on the optical response is an avenue that was explored, Sec.~\ref{sec:sensing}. In this work we have shown that the loss function as a function of the Fermi energy has an interesting dependence on the impurity type. 
Indeed the behaviour of the optical conductivity indicates that it may be possible to extract the value of the impurity strength, $\vimp$, and the 
impurity density, $\nimp$, and so extract the impurity resonance $\eres$. Finally the energy of the impurity position 
can be compared with the result from DFT simulations to determine the chemical compound present on the sample.  
The proposal for a sensor can take in account realistic laboratory constraints, so that it works 
at fixed incident light frequency and grating periodicity. This is accomplished using the tunability offered by graphene as a plasmonic material.

\begin{acknowledgments}
The authors wish to thank the Knut and Alice Wallenberg (KAW) foundation and the Swedish Foundation for Strategic Research (SSF) for 
financial support. We also thank Tomas L\"{o}fwander for stimulating discussions.
\end{acknowledgments}
%\bibliographystyle{apsrmp4-1}
\bibliography{FBA_mf_v1}

\end{document}